\documentclass[%
 reprint,
 amsmath,amssymb,
 aps,
]{revtex4-1}

\usepackage{graphicx}% Include figure files
\usepackage{dcolumn}% Align table columns on decimal point
\usepackage{bm}% bold math
\usepackage{amsmath}
\usepackage{graphicx}% Include figure files
\usepackage{dcolumn}% Align table columns on decimal point
\usepackage{bm}% bold math
\usepackage{amsmath}

\usepackage{polski}

\usepackage{epstopdf}
\usepackage{multirow}
\usepackage{siunitx}
\usepackage{color}
\usepackage{verbatim}
\usepackage{lipsum}
\usepackage[english]{babel}
\usepackage[T1]{fontenc}
\usepackage[utf8]{inputenc}
\usepackage[utf8]{inputenc}
\DeclareUnicodeCharacter{FB01}{fi}
\begin{document}

\preprint{APS/123-QED}

\title{Bi-stability of magnetic skyrmions in ultrathin multilayer nanodots induced by magnetostatic interaction}% Force line breaks with \\
% \thanks{, m.mru@amu.edu.pl}%

\author{M. Zelent$^{1}$}
\email{mateusz.zelent@amu.edu.pl}
\author{J. Tóbik$^{2}$}%
\author{M. Krawczyk$^{1}$}%
\author{K. Y. Guslienko$^{3,4}$}%
\author{M. Mruczkiewicz$^{2}$}%
 \email{m.mru@amu.edu.pl}
\affiliation{%
\textsuperscript{1}\,Faculty of Physics, Adam Mickiewicz University in Poznan, 61-614 Pozna\'{n}, Poland\\
  \textsuperscript{2}\,Institute of Electrical Engineering, Slovak Academy of Sciences, 841 04 Bratislava, Slovakia\\
  \textsuperscript{3}\,Depto. Fisica de Materiales, Universidad del País Vasco, UPV/EHU, 20018 San Sebastian, Spain \\
    \textsuperscript{4}\,IKERBASQUE, the Basque Foundation for Science, 48013 Bilbao, Spain
}%

% \collaboration{CLEO Collaboration}%\noaffiliation

\date{\today}% It is always \today, today,

\begin{abstract}
We report the results of simulations of magnetic skyrmion stability in ultrathin magnetic multilayer nanodots with interfacial Dzyaloshinskii-Moriya exchange interaction (DMI). We found that in presence of the lateral confinement the magnetostatic energy significantly influences the skyrmion stability and leads to stabilization of large radius skyrmion even at low values of the DMI strength, in addition to small radius skyrmion stabilized by DMI. In particular, stabilization of the skyrmion state with two different radii (bi-stability) is found in dipolarly-coupled (Pt/Co/Ir)$_{n}$ circular nanodots with the  number of repeats of the unit cell $n$ = 3 and 5. The bi-stability range is located at the DMI strength of 0.9-1.1 mJ/m$^2$ or at the total Co-layer thickness of 2.2-2.6 nm.
\end{abstract}

\pacs{Valid PACS appear here}% PACS, the Physics and Astronomy
\maketitle

%\tableofcontents

\section{\label{sec:level1}Introduction}
Magnetic skyrmions are topologically non-trivial inhomogeneous magnetization configurations that can be stabilized in thin ferromagnetic films or non-centrosymmetric bulk magnetic crystals. The strong spin orbit coupling and lack of the inversion symmetry give rise to the chiral Dzyaloshinskii-Moriya exchange interaction (DMI) \cite{117,118}, such as in bulk helimagnetic materials with B20 crystal lattice (bulk DMI), or at the interface between two dissimilar materials, ferromagnet/heavy metal (interfacial DMI) \cite{PhysRevLett.44.1538,Crepieux_1998_Dzyaloshinsky}.
Both types of DMI are considered to be favoring the chiral skyrmion stabilization, either in ground state (the lowest energy magnetic state) or metastable state (a higher energy local energy minimum). The bulk DMI leads to stability of a Bloch skyrmion, where magnetization rotates perpendicular to radial direction moving away from the center of the skyrmion and interfacial DMI stabilizes a N\'eel skyrmion (Fig. 1a), where magnetization rotates in the plane parallel to radial direction. Nanosize skyrmions have a potential to provide useful solutions for cheap low-power, high-density data storage and processing \cite{Beg_2015_Ground_state_search_hysteretic_behaviour,Yu_Room_temperature_skyrmion_shift_device,Fert_Skyrmions_on_the_track,Kiselev_chilar_skyrmions,Mruczkiewicz2017SpinStates}. The discovery of strong interfacial DMI in ultrathin Co/Pt, Co/Pd and Co/Pt/Ir multilayer films and dots expanded the possibilities for individual skyrmion stabilization in nanostructures  \cite{moreau2016additive,Pollard_2017_Observation}.

Moreau-Luchaire et al. \cite{moreau2016additive} observed by scanning transmission X-ray microscopy stable sub-\SI{100}{nm} magnetic skyrmions in dipolarly-coupled Pt/Co/Ir multilayers. Boulle et al. observed the N\'eel skyrmions in Pt/Co/MgO dots by photoemission electron microscopy combined with X-ray magnetic circular dichroism \cite{boulle2016room}. Recently, Pollard et al. \cite{Pollard_2017_Observation} reported on Lorentz microscopy imaging of less than \SI{100}{nm} stable skyrmions in exchange-coupled Co/Pd multilayers at room temperature. The skyrmion stability in infinite ferromagnetic film was studied by a semi-analytic approach and two types of skyrmions with different radii were found to be stable in some cases \cite{FButtner2017}.
It is expected, that lateral confinement modify the conditions for the skyrmion stability significantly \cite{Mruczkiewicz2017SpinStates,boulle2016room,Rohart2013_Skyrmion_confiment,moutafis2007magnetic,7061384,Zeissler2017PinningStacks}. 
\newline \indent Recent studies demonstrated, that the magnetic skyrmions are promising candidates for solving problems with unstable data storage, when  size of the memory unit cells is decreasing below a critical size resulting in data loss \cite{5,Yu2012SkyrmionDensity}. It was shown that the skyrmions with opposite core polarities reveal  hysteresis behavior when external magnetic field is varied \cite{Beg_2015_Ground_state_search_hysteretic_behaviour}. This feature is of interest to design a new generation of data storage devices. In Ref. \cite{Rohart2013_Skyrmion_confiment} the stability of a N\'eel skyrmion in circular nanodot was studied, and it was shown that due to the dot edge presence, the skyrmions with the sizes $R_{s}/R$ ($R_s$ is a skyrmion radius) between 0.50 and 0.78 are stable above critical value of DMI, $D_{C}$,  for the dot radii $R$=25-100 nm. In this study, the magnetostatic interaction was accounted as an effective uniaxial magnetic anisotropy. 

However, increasing the dot thickness the magnetostatic energy is expected to play important role in skyrmion stabilization even in ultrathin films/dots and should be taken into account \cite{boulle2016room,7061384}. The main two results of the present paper are as follows. Firstly, we show how the lateral confinement and corresponding magnetostatic energy lead to stabilization of the large radius skyrmions, if DMI strength is weak. Secondly, we show that due to the lateral confinement and strong magnetostatic interactions, the skyrmions can be bi-stable in multilayer circular dots even at zero magnetic field.
\begin{figure}[!ht]%
\centering
\includegraphics*[width=0.48\textwidth]{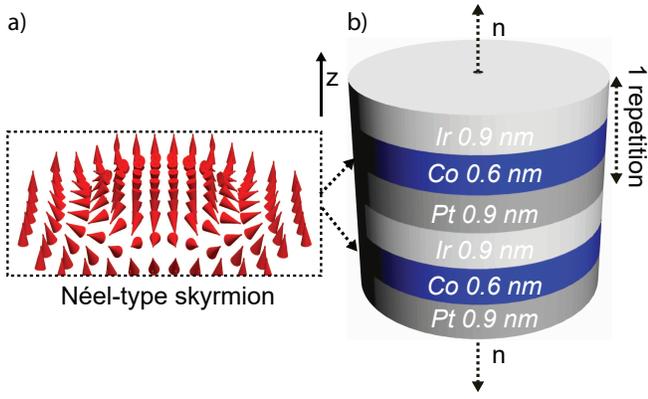}
\caption{%
a) Magnetization configuration of the ferromagnetic layers in the circular layered dot (N\'eel-type skyrmion). b) Sketch of the multilayer nanodot based on dipolarly coupled ultrathin Co layers with varied number of repeats, $n$, of the Ir/Co/Pt unit cell.
}\label{Fig:1}
\end{figure}
\section{Results}
A N\'eel skyrmion can be stabilized in a ultrathin dot within a finite range of interfacial DMI values, $D$. This range depends on competition of the DMI, exchange, anisotropy and magnetostatic interactions. The control of the magnetostatic interaction is realized by changing the number of the repeats, $n$, of the Co layers in the stack (Fig.~\ref{Fig:1}). For a description of the skyrmion we use the reduced magnetization components $ \boldsymbol {m}=\boldsymbol {M}/M_s = (m_r, m_{\phi}, m_z)$  in the cylindrical coordinate system with the $z$ axis directed along the dot thickness. 

\begin{figure}[!ht]
\includegraphics[width=0.48\textwidth]{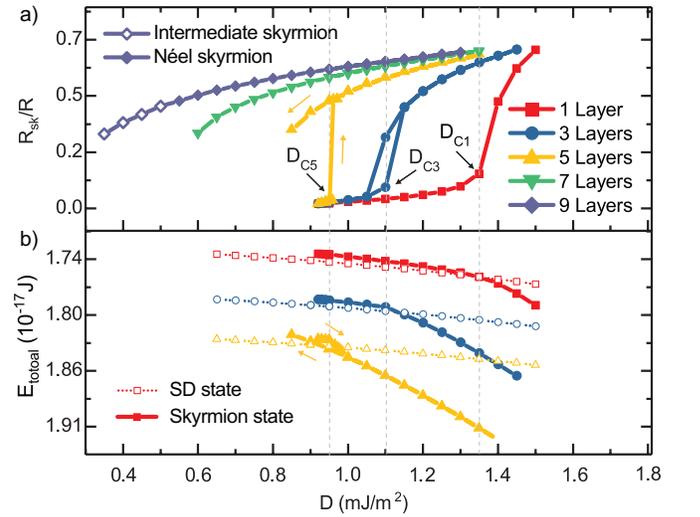}
\caption{ a) Normalized skyrmion radius, $R_{sk}$/$R$ as a function of DMI strength $D$ for different number of the ferromagnetic layers in the stack. $D_{C\textit{n}}$ represents the edge of stability of the small radius skyrmion and inflection point for rapid grow of skyrmion size with increase of DMI. b) Total magnetic energy (including the DMI, exchange, anisotropy and magnetostatic energies) for an isolated skyrmion (solid line), and for perpendicular single domain (SD) states as a function of DMI strength $D$ for different numbers of the ferromagnetic layers. }\label{Rs_od_dmi}
\end{figure}

Figure \ref{Rs_od_dmi} shows the dependence of normalized skyrmion radius $R_{sk}/R$ for $n=1-9$ repeats of the ferromagnetic layer of thickness, $t_{\textrm{Co}}=0.6$ nm, with Co-Co separation 1.8 nm, dot radius $R$=250 nm and magnetic parameters corresponding to Pt/Co/Ir dot given in Methods. The static skyrmion states were simulated with the  procedure described in Methods. All considered skyrmions possess the same core polarity, $m_{z}$ at the center of the dot, it is -1.

We can distinguish two types of skyrmions, small size and large size or bubble skyrmions, depending on DMI strength $D$ (Fig.~\ref{Rs_od_dmi}a). The clear border between the skyrmions cannot be defined and in some cases there is continuous transition between these two types of the skyrmions when DMI strength is varied, e.g, for $n=1$ (red line in Fig. \ref{Rs_od_dmi}a). We note, that there is strong influence of the number of ferromagnetic layer repeats on the large radius skyrmion stability and size, evidencing that the magnetostatic interaction plays an important role in stabilization of this skyrmion. The smaller skyrmion can nucleate at the DMI strength above $0.92$ mJ/m$^2$ and the range where it can exist decreases as the number of repeats increases. The points marked as $D_{C\textit{n}}$ indicate the edges of stability of the small radius skyrmions and inflation points for rapid grow of skyrmion size with increase of the DMI strength for each number $n$ of ferromagnetic layers in the stack. Increasing number of repeats decreases the critical DMI value where stable small radius skyrmion starts to grow. Thus, we conclude that the increase of magnetostatic interaction strength leads to decreasing the stability range of the small radius skyrmion.

Increasing number of repeats extends the range of DMI where large radius skyrmions can be nucleated.  However, the large DMI strength supports N\'eel-like skyrmion stabilization. Skyrmions with weak DMI strength lose the N\'eel-like character (the considerable $m_r$ magnetization component resulting in extra magnetostatic energy\cite{Guslienko_MMM_Magnetic_skyrmion_low_Frequency} cannot be compensated by DMI), intermediate states between Bloch- and N\'eel-like skyrmions appear, indicated with open dots in Fig. 2a \cite{Mruczkiewicz2017SpinStates}. Therefore, the hysteresis is a competition between small radius N\'eel-like skyrmions and large radius skyrmions having non-zero $m_{\phi}$ magnetization component (intermediate Bloch/N\'eel skyrmions \cite{Mruczkiewicz2017SpinStates}). Further increase of the number of ferromagnetic layer repeats and magnetostatic interaction strength in the layered stack leads to stabilization of the Bloch-like skyrmions even at zero DMI \cite{7061384}.

It was simulated (see Fig. 2a) that in some range of DMI and for certain number of the layer repeats, there is bi-stability of skyrmion with the same core polarity, but different sizes. For single layer ($n=1$) the bi-stability was not observed, however, it is present for 3 and 5 layers ($n=3, 5$). The difference between the large and small skyrmion radii, at the same DMI value, is relatively small for 3 layers (for $D=1.1$ mJ/m$^2$), but it increases for 5 layers (for $D=0.95$ mJ/m$^2$), and is \SI{75.5}{nm} and \SI{114.3}{nm}, respectively. Further increasing the number of repeats leads to destroying of small-radius N\'eel skyrmion stability. Therefore, the bi-stability can be observed only if magnetostatic interaction have intermediate strength competing with DMI. For 3 layers ($n=3$) we performed additional simulations, varying the dot radius $R$ between 100 and 1000 nm. We found that there is a critical value of $R$ that can support the skyrmion bi-stability and it is enhanced for large $R$. We also note that the range of $D$ where bi-stable skyrmions can exist expands with increasing the dot radius.

\begin{figure*}[htb]%
\centering
\includegraphics*[width=.78\textwidth,height=11cm]{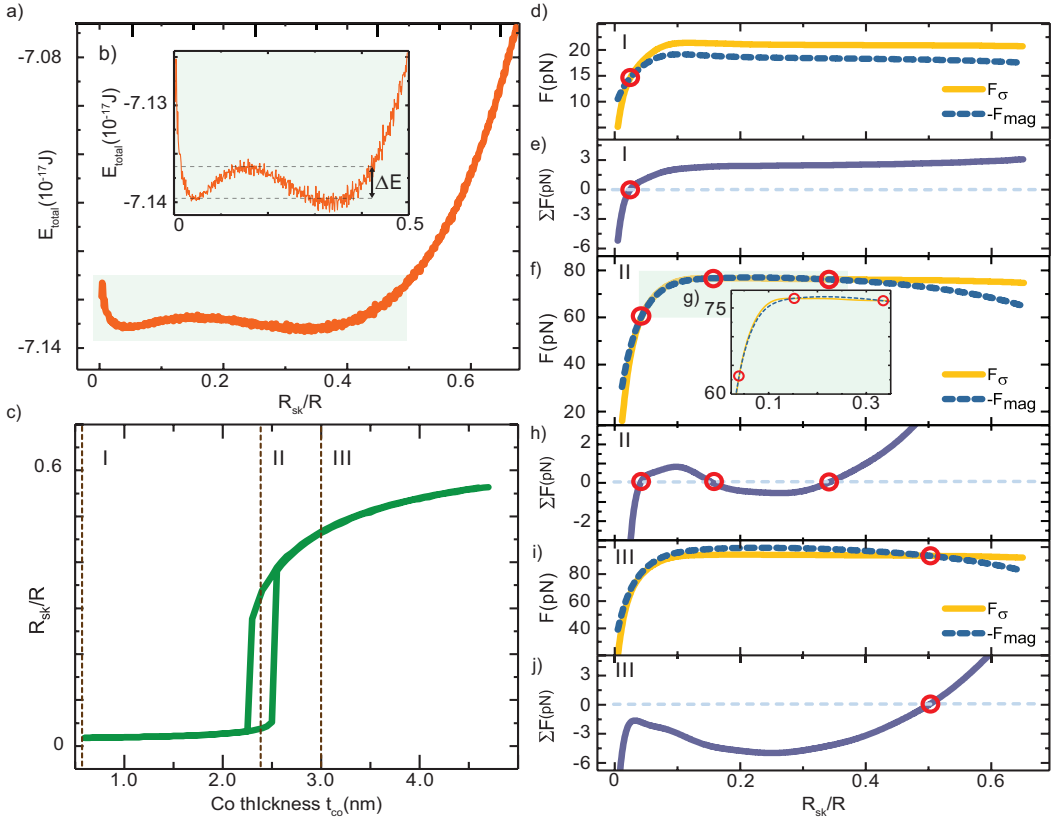}
\caption{The energy profile of skyrmion state as function of normalized skyrmion radius $R_{sk}$/$R$ in \SI{250}{nm} radius dot ($t_{Co}=2.38$ nm and $D$=0.95 mJ/m$^{2}$). b) Inset: enlarged energy profile with the indicated scale of total energy. c) Small and large skyrmion size as function of the thickness of Co layer. Dashed lines marked as I,II,III correspond to the force analysis presented in the panels (d-j). d,f,i) present the forces corresponding to the dot with different Co thicknesses, 0.6 nm (I), 2.38 nm (II) and 3.0 nm (III), respectively. $F_{mag}$ is a magnetostatic force, and remaining forces are presented by $F_{\sigma}$.  e,h,j) The sum of all forces is presented for I, II and III case, respectively, where zero value corresponds to the energy minimum. Red circles indicate energy minimum expressed as a crossing of the curves of partial forces. g) Inset: enlarged forces for (II).
}
\label{pochodne}
\end{figure*}

As shown in Fig. \ref{Rs_od_dmi}b the total magnetic energy (including the DMI, exchange,  anisotropy and magnetostatic energies) drawn as a solid line is higher than the single domain state energy for smaller skyrmion for each case. At higher values of $D$, the single domain state is no longer ground state, whereas the large radius skyrmion is. It is consistent with the previous observations \cite{sampaio2013nucleation}. 

\section{Forces analysis}
Stable skyrmions with different equilibrium radii can be characterized by their relative energies and energy  barrier separating them, see Fig. 3a. To investigate the origin of two minima in the energy profile, we have analyzed the total energy and forces (and their components) acting on the skyrmion when the radius is varied. In order to plot the energy profile as a function of skyrmion radius $E(R_{sk})$, we use the frozen spins technique (see Methods).  Since the DMI is strong enough to ensure the N\'eel type magnetization twisting, we assume that the frozen spins $m_{z}$ = 0, $m_{r}$ = 1  correspond to the lowest energy state at each radius of the ring, so the change of radius crosses the saddle point in energy function and allows us to estimate the minimum energy path and height of the energy barrier for bi-stable skyrmions.

We replace multilayer dot with simplified system, a single-layer circular dot with effective Co thickness, $t_{\textrm{Co}}$. That allows to continuously change the magnetostatic contribution to the total energy and study its influence on the skyrmion stabilization. Specific $t_{\textrm{Co}}$ could be also reproduced by a multilayer system with fitted $t_{\textrm{Co}}$ and Co-Co layer separation. Figure \ref{pochodne}c presents simulated sizes of stable skyrmion states in the nanodot as a function of the layer thickness, $t_{\textrm{Co}}$. Simulations showed that bi-stability exists and two possible skyrmion states are stable within the range of Co thickness 2.2--2.6 nm.

When the energy profile is calculated at $t_{\textrm{Co}} = \SI{0.6} {\nano\metre}$, single minimum is present around $R_{sk}=12$ nm ($R_{sk}/R=0.048$), see Figs. \ref{pochodne}d-e, and position of this minimum agrees with skyrmion stabilized without the use of frozen spins. The stabilization of this skyrmion is possible due to canceling of contributing forces: magnetostatic, exchange and anisotropy \cite{boulle2016room}. These forces are calculated from derivatives of contributions to the total energy with respect to the skyrmion radius $R_{sk}$ and are presented in Fig. \ref{pochodne}d as $F_{mag}$ for magnetostatic force and $F_{\sigma}$ for remaining forces. The sum of all forces $\Sigma F$ presented in Fig. \ref{pochodne}e) is zero at the energy minimum. The effect of the dot edge plays minor role on stabilization and this skyrmion state is also expected to exist in infinite film \cite{Rohart2013_Skyrmion_confiment}. The stabilization of this skyrmion is realized due to the nonlinear change in the magnetic energy as function of skyrmion radius induced when the skyrmion diameter is in order of domain wall (DW) width \cite{FButtner2017}.\footnote{In nanodot, this nonlinearity in energy as function of the skyrmion radius is also present when the difference between dot radius and skyrmion radius is in the range of domain wall width (or slightly higher due to spin canting at the dot boundaries). Thus, the stable skyrmions can be classified as small or large size in nanodot even when the magnetostatic interaction is taken into account as an effective easy-plane anisotropy \cite{Rohart2013_Skyrmion_confiment}}

When the Co thickness is increased to $t_{\textrm{Co}}$ = 2.38 nm, the skyrmions with two different sizes can be stable. The energy profile as function of the skyrmion radius is presented in Fig. 3a. The energy minima correspond to skyrmion sizes found in Fig. 3a at this value of $t_{\textrm{Co}}$. Analyzing the force $F_{mag}$ (Fig. 3f) one can observe significant curvature of the function at high values of $R_{sk}$. The curvature increases with  thickness (or $M_{s}$) increasing (Fig.  3i) and appears due to presence of the nanodot edge. The edge changes the $F_{mag}$ function and leads to stabilization of large radius skyrmion. Thus, presence of the lateral confinement modifies the magnetostatic energy introducing nonlinearity in the magnetostatic force $F_{mag}$ as function of the skyrmion radius (see Figs. 3f, 3i), and results in stabilization of skyrmion with size that is not related to the dot radius and DW width. The size of large skyrmion is not fixed to  $R_{sk}$/$R$ = 0.7-0.8 as in Ref. \cite{Rohart2013_Skyrmion_confiment}, but is proportional to the magnetostatic interaction strength. The Co thickness 6.6 nm ($n=11$) and $D=1.4-1.6$ mJ/m$^2$ \cite{moreau2016additive} correspond according to Figs. 2a and 3c to the stable N\'eel skyrmion with the radius $R_{sk}/R=0.6$.  

We distinguish two mechanisms of the skyrmion stabilization in nanodots: i) small radius N\'eel skyrmion is stabilized  by DMI when $R_{sk}$ is in range of the DW width, and ii) large radius skyrmion is stabilized by nonlinear increase of the magnetostatic interaction with skyrmion radius increasing. The latter is solely present in confined geometries. The stabilization of these skyrmions is independent and a bifurcation can be present in some cases, leading to the skyrmion bi-stability in nanodots.

Further increase of $t_{\textrm{Co}}$ leads to increase of curvature of the dependence $F_{mag}(R_{sk})$, crossing at small values of $R_{sk}$ is not possible and small radius skyrmion becomes unstable (Fig. 3i). The effect of the nonlinearity in the magnetostatic energy is very pronounced and large radius skyrmion is still stable.

An example shown in Fig. 3a for particular set of parameters exhibits a double well potential separated by small energy barrier, $\Delta E=2.33 \times 10^-20$ J. The energy barrier height can be enhanced and optimized via  control of the magnetic parameters, geometry or external magnetic field. The switching between two skyrmion states can be realized, e.g., by sweeping external magnetic field or applying a field pulse. Therefore, the skyrmion size hysteresis behavior can be realized without presence of a Bloch point as in switching between the skyrmions with opposite polarities \cite{Beg_2015_Ground_state_search_hysteretic_behaviour}.

%%%%%%%%%%%%%%%%%%%%%%%%%%%%%%%%%%%%
\section{Methods}
We performed finite-difference time-domain micromagnetic simulations with Mumax$^3$ solver \cite{mumax2011} using a uniformly discretized grid with the cell size 0.5-1.0 nm x 0.5-1.0 nm x 0.6-4.2 nm. We used a set of parameters, which describes a multilayer thin film with a perpendicular magnetic anisotropy and DMI induced at the interfaces. We took the parameters measured for multilayer Pt/Co/Ir dot in Ref. \cite{moreau2016additive}: saturation magnetization $M_{\text{s}} = 0.956$ MA/m, exchange stiffness $A=10.0$ pJ/m, and perpendicular magnetic anisotropy $K_{u} = 0.717$ MJ/m$^3$. They are distributed uniformly in 0.6 nm thick Co layer. Ultrathin Pt/Co/Ir dot total thickness below \SI{2.4} {\nano\metre} can be realized experimentally \cite{moreau2016additive,Pollard_2017_Observation}. To keep magnetostatic interactions between the layers we used air-gap separation instead of the nonmagnetic spacers since the interlayer exchange was neglected. Total thickness of one repeat was \SI{2.4}{nm}. In order to simulate multilayer stacks we used finite periodic boundary conditions \cite{mumax2011}. Magnetization non-uniformity through the thickness was neglected.
% * <kkingstoun@gmail.com> 2017-07-15T14:26:32.548Z:
%
% > The assumed value of the damping parameter taken into account in the FDTD simulations is $\alpha=0.01$, and it is close to the value of a Pt/Co/Ir.(potrzebne cytowanie).
%
%
% ^.
To overcome staircase effects resulting from the calculation method, we used built-in Mumax function \textit{edge smooth}.	
\newline \indent
The calculations of skyrmion size dependence on the DMI strength were performed by starting from an artificially nucleated skyrmion magnetic configuration, relaxing system and measuring the skyrmion diameter (if skyrmion was stable). We increased the interfacial DMI, equilibrated the system at every DMI step and used the previous magnetization configuration as an initial state in the subsequent DMI increase step. Next, the same procedure was repeated decreasing the DMI strength, starting from a large radius skyrmion at maximum DMI value possible.
\newline \indent
We used \textbf{frozen spins technique} to calculate the skyrmion magnetic energy  exploiting the Mumax build-in frozen spin function. The edge of the skyrmion ($m_{z}$=0, $m_{r}$=1) is defined as a narrow ring around the center of the nanodot. Such condition facilitates a N\'eel skyrmion configuration with radius corresponding to the radius of the ring. The rest of the spins are free and relax to a energy minimum. This method does not assume any shape of the skyrmion profile, only its radius, in contrast with the semi-analytical approach \cite{FButtner2017,7061384}. When varying the ring radius we obtain the total skyrmion energy as a function of its radius, see Fig. \ref{pochodne}a.

\section{Conclusions}
We analyzed influence of the magnetostatic interaction in circular multilayer nanodots on stability of the skyrmion magnetization configurations in ferromagnetic layers. We found that the skyrmions can be stabilized due to two different mechanisms, primary DMI or primary magnetostatic interaction leading to small and large size skyrmions, respectively. We found that these two kinds of the skyrmions can be stable simultaneously in the same nanodot. Thus, we demonstrated bi-stability of the skyrmion configurations with the same core polarities, but different sizes. The bi-stable skyrmions can be obtained in dots with realistic values of the dot sizes and the DMI strength, exchange and anisotropy parameters. Our results open a new route to design and develop a more efficient skyrmion memory, where information is coded as a skyrmion equilibrium size. The simulation approach based on the build-in Mumax function \textit{frozen spins} is able to estimate the minimum energy path between two N\'eel skyrmions with different radii.

\section{Acknowledgement}
The project is financed by SASPRO Program. The research was supported by the People Program (Marie  Curie  Actions) European Union's FP7 under REA grant agreement No. 609427 (project WEST 1244/02/01) and further co-funded by the Slovak Academy of Sciences and the European Union Horizon 2020 Research and Innovation Program under Marie Sklodowska-Curie grant agreement No.~644348. The financial assistance from the National Science Center of Poland (MagnoWa DEC-2-12/07/E/ST3/00538) is also acknowledged. K.G. acknowledges support by IKERBASQUE (the Basque Foundation for Science), and the Spanish MINECO grant FIS2016-78591-C3-3-R. The simulations were partially performed at the Poznan Supercomputing and Networking Center (Grant No.~209) and supported by the National Scholarship Program of the Slovak Republic funded by the Ministry of Education, Science, Research and Sport of the Slovak Republic.

\bibliography{Mendeley2}
\bibliographystyle{apsrev4-1}
\end{document}